# Methylacetylene ($CH_3CCH$) and propene ($C_3H_6$) formation in cold dense clouds: a case of dust grain chemistry.

*Kevin M. Hickson*[1,2], *Valentine Wakelam*[3,4], *Jean-Christophe Loison*[1,2]*

*Corresponding author: jean-christophe.loison@u-bordeaux.fr

[1] *Univ. Bordeaux, ISM, UMR 5255, F-33400 Talence, France*

[2] *CNRS, ISM, UMR 5255, F-33400 Talence, France*

[3] *Univ. Bordeaux, LAB, UMR 5804, F-33270, Floirac, France.*

[4] *CNRS, LAB, UMR 5804, F-33270, Floirac, France*

We present an extensive review of gas phase reactions producing methylacetylene and propene showing that these relatively abundant unsaturated hydrocarbons cannot be synthesized through gas-phase reactions. We explain the formation of propene and methylacetylene through surface hydrogenation of $C_3$ depleted onto interstellar ices, $C_3$ being a very abundant species in the gas phase.

## 1. INTRODUCTION

The formation of Complex Organic Molecules (COMs) in the interstellar medium is a challenging issue, specifically with respect to their observation in dense molecular clouds. This is particularly true for propene ($CH_3$-CH=$CH_2$, hereafter named $C_3H_6$) and methylacetylene ($CH_3$-CCH), their formation in dense molecular clouds is not well understood. Gas phase formation routes for these compounds have been extensively investigated but fail to reproduce the observed abundances of these species in dense molecular clouds (Lin *et al*. 2013, Öberg *et al*. 2013, Vastel *et al*. 2014) suggesting that important formation pathways could be missing at low temperature.

Methylacetylene was first detected in TMC-1 35 years ago (Irvine *et al.* 1981) with an abundance of around $6.10^{-9}$ relative to $H_2$. Askne et al (1983) and Kuiper et al (1984) detected methylacetylene in various interstellar media (DR21(OH), TMC-1, W3(OH), W51, Orion A, Sgr B2, ..) with a variable abundance around a few $10^{-9}$ relative to $H_2$. The authors of both of these papers proposed that methylacetylene could be used as a probe of the kinetic temperature of interstellar clouds. Methylacetylene has also been detected in small translucent clouds (Turner *et al.* 2000, Turner *et al.* 1999) and in star forming regions (Miettinen *et al.* 2006), with a positive correlation between the fractional $CH_3CCH$ abundance and $T_{kin}$, suggesting a possible desorption scenario with grain surface production.

The first positive detection of propene was much more recent, firstly in TMC-1 (Marcelino *et al.* 2007) and then in four additional cold dense clouds Lupus-1A, L1495B, L1521F and Serpens South (Agúndez *et al.* 2015) with an average abundance around $2 \cdot 10^{-9}$ relative to $H_2$.

In this study, we review the reactions producing methylacetylene and propene. Firstly, we review the reactions present in the usual astrochemical databases (KIDA (Wakelam *et al.* 2015) and UMIST (McElroy *et al.* 2012)). Then we perform an extensive search for new gas phase reactions leading to these unsaturated hydrocarbons. In the end, to explain the observation of propene and methylacetylene we propose efficient surface synthesis followed by a release into the gas phase through chemical desorption mechanisms (Vasyunin & Herbst 2013, Garrod *et al.* 2007). We also focus on the critical role of the O + $C_3$ reaction.

## 2. The chemical model

To calculate the methylacetylene and propene abundances we use the chemical model Nautilus (Hersant *et al.* 2009, Semenov *et al.* 2010). The Nautilus code computes the gas-phase and grain surface composition as a function of time taking into account reactions in the gas-phase, physisorption onto and desorption from grain surfaces and reactions at the surface. The surface reactions are based on the Langmuir-Hinshelwood mechanism with the formalism of Hasegawa *et al.* (1992) and includes the low temperature Eley-Rideal and complexation mechanisms considered by Ruaud et al (2015). For desorption, we consider thermal desorption as well as desorption induced by cosmic-rays (Hasegawa & Herbst 1993) and chemical reactions (the exothermicity of surface chemical reactions allows for the species to be desorbed after their formation) (Garrod et al. 2007). We consider in the nominal model that 1% of the newly formed species desorb and 99% remain on the grain surfaces.

The gas-phase network is based on kida.uva.2014 (http://kida.obs.u-bordeaux1.fr/models (Wakelam et al. 2015)), with the modifications described in section 3. The surface network and parameters are the same as in (Ruaud et al. 2015). Following (Hincelin *et al.* 2015), the encounter desorption mechanism is included in the code. This mechanism accounts for the fact that the $H_2$ binding energy on itself is much smaller than on water ices and prevents the formation of several $H_2$ monolayers on grain surfaces.

The chemical composition of the gas-phase and the grain surfaces is computed as a function of time. The gas and dust temperatures are equal to 10 K, the total H density is equal to $2{\cdot}10^4$ cm$^{-3}$ for molecular clouds. The cosmic-ray ionization rate is equal to $1.3{\cdot}10^{-17}$ s$^{-1}$ and the visual extinction is equal to 30. All elements are assumed to be initially in atomic form, except for hydrogen, which is entirely molecular. The initial abundances are similar to those of Table 1 of Hincelin *et al.* (2011), the C/O elemental ratio being equal to 0.7 in this study. We also considered larger C/O elemental ratios up to 1.0 but found that the $C_3H_4$ and $C_3H_6$ abundances are not sensitive to this parameter (propene reacts relatively quickly with oxygen atoms but also with carbon atoms. On the other hand, the high carbon abundance favours hydrocarbon formation). The grains are considered to be spherical with a 0.1μm radius, a 3 g.cm$^{-3}$ density and about $10^6$ surface sites, all chemically active. The dust to gas mass ratio is set to 0.01 and the barrier to diffusion for all species is assumed to be half of their binding energies.

## 3. Review of the gas phase reactions

We performed an extensive review of potential gas phase chemical schemes leading to methylacetylene and propene. A large part, but not all, of these reactions were already included in current networks, but for all of them we checked the cited references leading to various changes in the rate constants and branching ratios. The important reactions for methylacetylene and propene production and loss in the conditions of cold interstellar clouds are listed in Table I including an extensive bibliography presented in the supplementary information.

The main gas phase source of $CH_3CCH$ is thought to be ionic reactions producing $C_3H_5^+$ followed by dissociative recombination (DR) with minor contributions from neutral reactions, mainly $CH + C_2H_4 \rightarrow CH_3CCH + H$ (the branching ratio of this reaction toward methylacetylene and/or allene production is subject to discussion (Goulay *et al.* 2009, Zhang *et al.* 2012)). The most efficient source of methylacetylene through $C_3H_5^+$ production was thought to be the $C_3H_3^+ + H_2 \rightarrow C_3H_5^+ + h\nu$ (1) radiative association reaction. Herbst et al (2010) performed theoretical calculations leading to a high rate constant for this process, in disagreement with earlier experimental results on the analogous three-body association process measured by (McEwan *et al*. 1999) which was found to be very slow. Some of these authors revisited reaction (1) some years later (Lin et al. 2013) through a coupled experimental and theoretical study showing that, no matter which of the isomers of $C_3H_3^+$ are considered, the radiative association leading to protonated methylacetylene has sufficiently high barriers that these reactions are unimportant in the cold interstellar medium and cannot be an efficient way to form methylacetylene in molecular clouds. Other gas phase reactions can produce $C_3H_5^+$, and then methylacetylene through DR, the most efficient being $C_2H_3^+ + CH_4 \rightarrow C_3H_5^+ + H_2$ (Anicich 2003) with secondary contributions from $C_2H_2^+ + CH_4$, $CH_3^+ + C_2H_4$, $C_2H_4^+ + C_2H_4$ as well as $C^+ + C_3H_6$ (Anicich 2003). Nevertheless, all of these ionic reactions involve fluxes that are too small to produce the observed methylacetylene abundances. It should be noted that the protonation reactions of methylacetylene ($CH_3CCH + H_3^+$, $HCO^+$ ... ) lead to $C_3H_5^+$, but these processes act as destruction mechanisms for $CH_3CCH$ as the dissociative recombination (DR) of $C_3H_5^+$ does not lead back to $CH_3CCH$ with 100% efficiency. With regard to neutral reactions, the $CH + C_2H_4 \rightarrow CH_3CCH + H$ reaction produces directly some methylacetylene with a large rate constant at low temperature but this process involves small fluxes. The $C + C_2H_5 \rightarrow CH_3CCH + H$ and $H + C_3H_5 \rightarrow CH_3CCH + H_2$ reactions also involve small fluxes considering only gas-phase networks as $C_2H_5$ and $C_3H_5$ are essentially formed on grains, or through the protonation of $C_2H_6$ and $C_3H_6$ (both essentially formed on grains) followed by DR of $C_2H_7^+$ and $C_3H_7^+$. Then the gas phase abundances of $C_2H_5$ and $C_3H_5$ are strongly related to grain chemistry so that gas phase reactions alone do not produce large amounts of $C_2H_5$ and $C_3H_5$. These gas phase reactions alone are not sufficient to reproduce the observations of methylacetylene in dense molecular clouds. As shown in Fig. 1 (continuous line), the gas phase formation of $CH_3CCH$ leads to abundances three orders of magnitude below the observations even with the reaction review listed in Table I.

**Table I: Summary of reaction review.**

$k = \alpha\cdot(T/300)^{\beta}\cdot\exp(-\gamma/T)$ cm$^3$ molecule$^{-1}$ s$^{-1}$ , T range is 10-300K
except for :
Ionpol1: $k = \alpha\beta(0.62+0.4767\gamma\ (300/T)^{0.5})$ cm$^3$ molecule$^{-1}$ s$^{-1}$ , (Wakelam *et al.* 2012, Wakelam *et al.* 2010)
Ionpol2: $k = \alpha\beta(1+0.0967\gamma\ (300/T)^{0.5}+(\gamma^2/10.526)\ (300/T))$ cm$^3$ molecule$^{-1}$ s$^{-1}$ ,
$F_0 = \exp(\Delta k/k_0)$ and $F(T)=F_0\cdot\exp(g\cdot|1/T-1/T_0|)$

| | **Reaction** | **α** | **β** | **γ** | **$F_0$** | **g** | **ref** |
|---|---|---|---|---|---|---|---|
| 1. | $H + C_3H_5$ → $H_3C\text{-}CCH + H_2$<br>→ $C_3H_6 + h\nu$ | 1.5e-11<br>9.2e-13 | 0<br>-1.5 | 0<br>0 | 3<br>10 | 0<br>0 | (Hanning-Lee & Pilling 1992, Harding *et al.* 2007, Hébrard *et al.* 2013). H abstraction may be very low at 10K. |
| 2. | $C^+ + C_3H_4$ → $C_4H_2^+ + H_2$<br>→ $C_3H_4^+ + C$<br>→ $C_3H_3^+ + CH$<br>→ $C_2H_2^+ + C_2H_2$<br>→ $C_2H_3^+ + C_2H$ | 5.7e-10<br>5.7e-10<br>3.8e-10<br>1.9e-10<br>1.9e-10 | -0.5<br>-0.5<br>-0.5<br>-0.5<br>-0.5 | 0<br>0<br>0<br>0<br>0 | 1.6<br>1.6<br>1.6<br>1.6<br>1.6 | 0<br>0<br>0<br>0<br>0 | (Bohme *et al.* 1982) considering the dipole of $C_3H_4$ |
| 3. | $C^+ + C_3H_6$ → $C_2H_3^+ + C_2H_3$<br>→ $C_3H_5^+ + CH$<br>→ c-$C_3H_3^+ + CH_3$<br>→ $C_2H_2^+ + C_2H_4$<br>→ $C_3H_6^+ + C$<br>→ $C_4H_3^+ + H_2 + H$ | 6.0e-10<br>4.0e-10<br>3.0e-10<br>3.0e-10<br>2.0e-10<br>2.0e-10 | -0.5<br>-0.5<br>-0.5<br>-0.5<br>-0.5<br>-0.5 | 0<br>0<br>0<br>0<br>0<br>0 | 1.6<br>1.6<br>1.6<br>1.6<br>1.6<br>1.6 | 0<br>0<br>0<br>0<br>0<br>0 | (Bohme et al. 1982) considering the dipole of $C_3H_4$ |
| 4. | $C^+ + C_3H_8$ → $C_2H_5^+ + C_2H_3$<br>→ $C_2H_3^+ + C_2H_5$<br>→ $C_2H_2^+ + C_2H_6$ | 7.0e-10<br>5.0e-10<br>4.0e-10 | 0<br>0<br>0 | 0<br>0<br>0 | 1.6<br>1.6<br>1.6 | 0<br>0<br>0 | (Bohme et al. 1982) |
| 5. | C + $C_2H_5$ → $CH_3CCH$ + H<br>→ $CH_2CCH_2$ + H | 1.0e-10<br>1.0e-10 | 0<br>0 | 0<br>0 | 3<br>0 | 0<br>0 | KIDA, close to capture rate. |
| 6. | C + $C_3H_5$ → H + $CH_2CHC_2H$ | 2.0e-10 | 0 | 0 | 4 | 0 | Rate constant from capture rate theory and branching ratio deduced from theoretical study of the isomeric CH + $C_3H_4$ reaction (Goulay et al. 2009, Loison & Bergeat 2009). |
| 7. | C + $C_3H_6$ → $CH_3$ + $C_3H_3$<br>→ H + $C_4H_5$ | 2.0e-10<br>1.0e-10 | 0<br>0 | 0<br>0 | 1.6<br>1.6 | 0<br>0 | (Chastaing *et al.* 1999, Loison & Bergeat 2004, Chin *et al.* 2013) |
| 8. | C + $C_3H_7$ → H + $CH_2CHCHCH_2$<br>→ $CH_3$ + $CH_2CHC_2H$ | 1.6e-10<br>4.0e-11 | 0<br>0 | 0<br>0 | 4<br>4 | 0<br>0 | Rate constant from capture rate theory and branching ratio deduced from theoretical study of the isomeric CH + $C_3H_6$ (Trevitt *et al.* 2013, Loison & Bergeat 2009). |
| 9. | CH + $C_2H_4$ → $H_2CCCH_2$ + H<br>→ $CH_3CCH$ + H | 2.4e-10<br>1.0e-10 | -0.546<br>-0.546 | 29.6<br>29.6 | 1.6<br>1.6 | 7<br>7 | (Butler *et al.* 1981, Berman *et al.* 1982, Thiesemann *et al.* 1997, Thiesemann *et al.* 2001, Canosa *et al.* 1997, McKee *et al.* 2003, Goulay et al. 2009, Gosavi *et al.* 1985, Wang & Huang 1998, Davis *et al.* 1999, Stranges *et al.* 2008, Zhang et al. 2012, Loison & Bergeat 2009) |
| 10. | CH + $C_2H_6$ → $CH_3$ + $C_2H_4$<br>→ H + $C_3H_6$ | 2.8e-11<br>5.0e-12 | -0.648<br>-0.648 | 43.6<br>43.6 | 1.6<br>1.6 | 7<br>7 | (Butler *et al.* 1980, Butler et al. 1981, Berman & Lin 1983, Canosa et al. 1997, McKee et al. 2003, Galland *et al.* 2003) |
| 11. | CH + $C_3H_4$ → H + $CH_2CHC_2H$ | 4.2e-10 | 0 | 0 | 1.4 | 14 | (Fleming *et al.* 1980, Butler et al. 1981, Daugey *et al.* 2005, Goulay et al. 2009, Loison & Bergeat 2009) |
| 12. | CH + $C_3H_6$ → H + $CH_2CHCHCH_2$<br>→ $CH_3$ + $CH_2CHC_2H$ | 3.3e-10<br>9.0e-11 | 0<br>0 | 0<br>0 | 1.4<br>1.4 | 7<br>7 | (Daugey et al. 2005, Trevitt et al. 2013, Loison & Bergeat 2009) |
| 13. | $CH_3^+$ + $C_2H_4$ → $C_2H_3^+$ + $CH_4$<br>→ $C_3H_3^+$ + $H_2$ + $H_2$<br>→ $C_3H_5^+$ + $H_2$ | 4.9e-10<br>4.0e-11<br>5.4e-10 | 0<br>0<br>0 | 0<br>0<br>0 | 1.2<br>1.2<br>1.2 | 0<br>0<br>0 | (Anicich 2003) |
| 14. | $CH_3^+$ + $C_2H_6$ → $C_2H_5^+$ + $CH_4$<br>→ $C_3H_5^+$ + $H_2$ + $H_2$<br>→ $C_3H_7^+$ + $H_2$ | 1.5e-9<br>1.1e-10<br>8.0e-11 | 0<br>0<br>0 | 0<br>0<br>0 | 1.6<br>1.6<br>1.6 | 0<br>0<br>0 | (Anicich 2003) |
| 15. | $CH_3^+$ + $C_3H_8$ → $C_3H_7^+$ + $CH_4$ | 1.0e-9 | 0 | 0 | 1.6 | 0 | (Anicich 2003) and by comparison with $CH_3^+$ + $C_2H_6$ |
| 16. | $CH_3$ + $C_2H_3$ → $C_3H_5$ + H | 1.0e-10<br>3.3e-11 | 0<br>0 | 0<br>0 | 1.4<br>1.4 | 0<br>0 | At low pressure (Fahr *et al.* 1999, Stoliarov *et al.* 2000, Hébrard et al. 2013, Thorn *et al.* 2000) |
| | → $CH_4 + C_2H_2$ | | | | | | |
| 17. | $C_2H + C_3H_4$ → $C_5H_4 + H$ | 2.1e-10 | -0.3 | 0 | 1.6 | 0 | (Carty *et al.* 2001) |
| 18. | $C_2H + C_3H_6$ → $C_4H_4 + CH_3$ | 2.0e-10 | 0 | 0 | 1.3 | 0 | (Chastaing *et al.* 1998, Vakhtin *et al.* 2001, Woon & Park 2009) |
| 19. | $C_2H_2^+ + CH_4$ → $C_3H_4^+ + H_2$<br>→ $C_3H_5^+ + H$ | 1.9e-10<br>7.0e-10 | 0<br>0 | 0<br>0 | 1.2<br>1.2 | 0<br>0 | (Anicich 2003) |
| 20. | $C_2H_3^+ + CH_4$ → $C_3H_5^+ + H_2$ | 1.9e-10 | 0 | 0 | 1.6 | 0 | (Anicich 2003) |
| 21. | $C_2H_4^+ + C_2H_2$ → $C_3H_3^+ + CH_3$<br>→ $C_4H_5^+ + H$ | 6.5e-10<br>1.9e-10 | 0<br>0 | 0<br>0 | 1.2<br>1.2 | 0<br>0 | (Anicich 2003) |
| 22. | $C_2H_4^+ + C_2H_4$ → $C_3H_5^+ + CH_3$<br>→ $C_4H_7^+ + H$ | 7.2e-10<br>7.1e-11 | 0<br>0 | 0<br>0 | 1.2<br>1.2 | 0<br>0 | (Anicich 2003) |
| 23. | $C_2H_5^+ + CH_4$ → $C_3H_7^+ + H_2$ | 6.0e-13 | 0 | 1260 | 1.6 | 400 | (Hiraoka & Kebarle 1975, Hiraoka & Kebarle 1976, Hiraoka *et al.* 1993, Collins & O'Malley 1994) |
| 24. | $C_2H_4^+ + C_2H_6$ → $C_3H_6^+ + CH_4$<br>→ $C_3H_7^+ + CH_3$ | 3.6e-13<br>4.8e-12 | 0<br>0 | 0<br>0 | 1.2<br>1.2 | 100<br>100 | (Anicich 2003) |
| 25. | $C_2H_5^+ + C_2H_6$ → $C_3H_7^+ + CH_4$<br>→ $C_4H_9^+ + H_2$ | 5.46e-12<br>3.35e-11 | 0<br>0 | 0<br>0 | 1.2<br>1.20 | 0<br>0 | (Anicich 2003) |
| 26. | $C_3H_6 + H_3^+$ → $C_3H_7^+ + H_2$ | 1.0 | 3.5e9 | 0.51 | 3 | 0 | Ionpol2 |
| 27. | $C_3H_6 + HCO^+$ → $C_3H_7^+ + CO$ | 1.0 | 1.42e-9 | 0.51 | 3 | 0 | Ionpol2 |
| 28. | $C_3H_6 + H_3O^+$ → $C_3H_7^+ + H_2O$ | 1.0 | 1.65e-9 | 0.51 | 3 | 0 | Ionpol2 |
| 29. | $C_3H_6 + HCNH^+$ → $C_3H_7^+ + HCN$ | 1.0 | 1.43e-9 | 0.51 | 3 | 0 | Ionpol2 |
| 30. | $C_3H_8 + H_3^+$ → $C_3H_9^+ + H_2$ | 3.31e-9 | 0 | 0 | 3 | 0 | Capture rate |
| 31. | $C_3H_8 + HCO^+$ → $C_3H_9^+ + CO$ | 1.33e-9 | 0 | 0 | 3 | 0 | Capture rate |
| 32. | $C_3H_8 + H_3O^+$ → $C_3H_9^+ + H_2O$ | 1.52e-9 | 0 | 0 | 3 | 0 | Capture rate |
| 33. | $O + C_3H_5$ → $C_2H_3CHO + H$<br>→ $C_2H_4 + H + CO$<br>→ $H_2CO + C_2H_2 + H$ | 9.0e-11<br>8.0e-11<br>1.0e-11 | 0<br>0<br>0 | 0<br>0<br>0 | 1.6<br>1.6<br>1.6 | 0<br>0<br>0 | (Hoyermann *et al.* 2009) |
| 34. | $O + C_3H_6$ → $C_2H_5 + H$ CO<br>→ $CH_3 + CH_2$ CHO | 3.6e-12<br>4.4e-12 | 0<br>0 | 0<br>0 | 3<br>3 | 0<br>0 | Rate constant in the 10-50K deduced from (Sabbah *et al.* 2007), branching ratio mainly from (Savee *et al.* 2012) (see also (Knyazev *et al.* 1992, Cavallotti *et al.* 2014, Leonori *et al.* 2015)) |
| 35. | $O + C_3H_7$ → $C_2H_5CHO + H$<br>→ $H_2CO + C_2H_5$<br>→ $CH_3C(O)CH_3 + H$ | 6.0e-11<br>2.0e-11<br>4.0e-11<br>4.0e-11 | 0<br>0<br>0<br>0 | 0<br>0<br>0<br>0 | 1.8<br>1.8<br>1.8<br>1.8 | 0<br>0<br>0<br>0 | (Tsang & Hampson 1986) with branching ratio from (Hoyermann & Sievert 1979) considering similar amount of i-$C_3H_7$ and n-$C_3H_7$. |
| | → $CH_3CHO + CH_3$ | | | | | | |
| 36. | $N + C_3H_5$ → $C_2H_3CN + H_2$<br>→ $C_2H_4 + HCN$<br>→ $NH + C_3H_4$ | 3.2e-11<br>3.2e-11<br>1.3e-11 | 0.17<br>0.17<br>0.17 | 0<br>0<br>0 | 3<br>3<br>2 | 0<br>0<br>0 | By comparison with $N + C_2H_3$. We consider that $^2CH_2$=CH-CHN· react with H atom on singlet surface leading to $^1C_2H_3CN + {}^1H_2$ and $^1C_2H_4 + {}^1HCN$. |
| 37. | $N + C_3H_7$ → $C_2H_5 + H_2CN$ | 1.0e-10 | 0 | 0 | 4 | 0 | By comparison with $N+C_2H_5$ considering only n-$C_3H_7$ |
| | | | | | | | |
| 38. | $C_3H_4^+ + e^-$ → $H + C_3H_3$<br>→ H + H + c-$C_3H_2$<br>→ H + H + l-$C_3H_2$<br>→ H + H + t-$C_3H_2$<br>→ $CH_3 + C_2H$ | 3.0e-7<br>1.0e-7<br>1.0e-7<br>1.0e-7<br>1.0e-7 | -0.7<br>-0.7<br>-0.7<br>-0.7<br>-0.7 | 0<br>0<br>0<br>0<br>0 | 2<br>2<br>2<br>2<br>2 | 0<br>0<br>0<br>0<br>0 | Rate constant from (Florescu-Mitchell & Mitchell 2006), 89.7 % de $C_3H_x$ (supposed to be $C_3H_4$**) and 10.3% of $CH_x + C_2H_y$ (Angelova *et al.* 2004a). We supposed than $C_3H_3$** leads mainly to $C_3H_2$ by comparison with $C_3H_3$ photodissociation (Deyerl *et al.* 1999). |
| 39. | $C_3H_5^+ + e^-$ → H + $CH_3CCH$<br>→ H + $CH_2CCH_2$<br>→ H + H + $C_3H_3$<br>→ $H_2 + C_3H_3$<br>→ $H + H_2$ + c-$C_3H_2$<br>→ $H + H_2$ + l-$C_3H_2$<br>→ $H + H_2$ + t-$C_3H_2$<br>→ $CH_3 + C_2H_2$<br>→ $H + CH_2 + C_2H_2$ | 7.0e-8<br>7.0e-8<br>1.1e-7<br>4.2e-8<br>1.0e-7<br>5.0e-8<br>5.0e-8<br>4.7e-8<br>4.7e-8 | -0.7<br>-0.7<br>-0.7<br>-0.7<br>-0.7<br>-0.7<br>-0.7<br>-0.7<br>-0.7 | 0<br>0<br>0<br>0<br>0<br>0<br>0<br>0<br>0 | 2<br>2<br>2<br>2<br>2<br>2<br>2<br>2<br>2 | 0<br>0<br>0<br>0<br>0<br>0<br>0<br>0<br>0 | 86.7 % de $C_3H_x$ (supposed to be $C_3H_4$**) and 13.3% of $CH_x + C_2H_y$ (Angelova et al. 2004a, Angelova *et al.* 2004b). We supposed than $C_3H_4$** leads mainly to $C_3H_3$ by comparison with $C_3H_4$ photodissociation (Seki & Okabe 1992, Ni *et al.* 1999, Harich *et al.* 2000a, Harich *et al.* 2000b). |
| 40. | $C_3H_7^+ + e^-$ → $H + C_3H_6$<br>→ H + H + $C_3H_5$ | 3.4e-7<br>8.8e-8<br>3.6e-8<br>3.6e-8<br>1.52e-7 | -0.7<br>-0.7<br>-0.7<br>-0.7<br>-0.7 | 0<br>0<br>0<br>0<br>0 | 2<br>2<br>2<br>2<br>2 | 0<br>0<br>0<br>0<br>0 | $k = 8e\text{-}7*(T/300)^{-0.7}$ (Florescu-Mitchell & Mitchell 2006), branching ratio deduced from (Janev & Reiter 2004, Reiter & Janev 2010, Angelova et al. 2004b) |
| | → H + $H_2$ + $CH_3CCH$<br>→ H + $H_2$ + $CH_2CCH_2$<br>→ H + $CH_3$ + $C_2H_3$<br>→ $CH_3$ + $C_2H_4$<br>→ $CH_3$ + $C_2H_2$ + $H_2$ | 3.2e-8<br>8.8e-8 | -0.7<br>-0.7 | 0<br>0 | 2<br>2 | 0<br>0 | |
| 41. | $C_3H_9^+ + e^-$ → H + $C_3H_8$<br>→ H + H + $C_3H_7$<br>→ H + $H_2$ + $C_3H_6$<br>→ $CH_3$ + $C_2H_6$<br>→ $CH_3$ + $C_2H_4$ + $H_2$ | 2.0e-7<br>2.0e-7<br>2.0e-7<br>1.0e-7<br>1.0e-7 | -0.7<br>-0.7<br>-0.7<br>-0.7<br>-0.7 | 0<br>0<br>0<br>0<br>0 | 2<br>2<br>2<br>2<br>2 | 0<br>0<br>0<br>0<br>0 | Deduced from $C_3H_7^+ + e^-$ |
| | | | | | | | |
| 42. | s-H + s-$C_3$ → s-l-$C_3H$<br>→ s-c-$C_3H$ | 0.5<br>0.5 | | 0<br>0 | | 0<br>0 | (Mebel & Kaiser 2002) |
| 43. | s-H + s-l-$C_3H$ → s-l-$C_3H_2$<br>→ s-c-$C_3H_2$ | 0.5<br>0.5 | | 0<br>0 | | 0<br>0 | (Mebel *et al.* 1998) |
| 44. | s-H + s-c-$C_3H$ → s-l-$C_3H_2$<br>→ s-c-$C_3H_2$ | 0.5<br>0.5 | | 0<br>0 | | 0<br>0 | (Mebel et al. 1998) |
| 45. | s-H + s-t,c-$C_3H_2$ → s-$C_3H_3$ (s-$CH_2CCH$) | 1.0 | | 0 | | 0 | (Nguyen *et al.* 2001) |
| 46. | s-H + s-$C_3H_3$ → s-$CH_3CCH$ | 1.0 | | 0 | | 0 | (Miller & Klippenstein 2003) |
| 47. | s-H + s-$CH_3CCH$ → s-$C_3H_5$ | 1.0 | | 2013 | | 1000 | (Faravelli *et al.* 2000, Miller *et al.* 2008) |
| 48. | s-H + s-$C_3H_5$ → s-$C_3H_6$ | 1.0 | | 0 | | 0 | (Qu *et al.* 2009) |
| 49. | s-H + s-$C_3H_6$ → s-$C_3H_7$ | 1.0 | | 1600 | | 1000 | (Tsang 1991, Seakins *et al.* 1993, Galland et al. 2003) |
| 50. | s-H + s-$C_3H_7$ → s-$C_3H_8$ | 1.0 | | 0 | | 0 | (Harding *et al.* 2005) |
| 51. | s-H + s-$C_3H_8$ → s-$C_3H_7$ + s-$H_2$ | 1.0 | | 4000 | | 1000 | NIST |
| 52. | s-$H_2$ + s-$C_3$ → s-l-$C_3H$ + s-H<br>→ s-c-$C_3H$ + s-H | 0<br>0 | | | | | Widely endothermic<br>Widely endothermic |
| 53. | s-$H_2$ + s-l-$C_3H$ → s-l-$C_3H_2$ + s-H<br>→ s-c-$C_3H_2$ + s-H | 0<br>0 | | | | | Widely endothermic<br>Large barrier (this work) |
| | → s-$C_3H_3$ | 1.0 | | 3600 | | | M06-2X/cc-pVTZ/TST calculations (this work) |
| 54. | s-$H_2$ + s-c-$C_3H$ → s-c-$C_3H_2$ + s-H<br>→ s-$C_3H_3$ | 1.0<br>0 | | 6700 | | | M06-2X/cc-pVTZ/TST calculations (this work) |
| 55. | s-$H_2$ + s-l-$C_3H_2$ → s-$C_3H_3$ + s-H | 0 | | | | | endothermic |
| 56. | s-$H_2$ + s-c-$C_3H_2$ → s-$C_3H_3$ + s-H | 0 | | | | | endothermic |

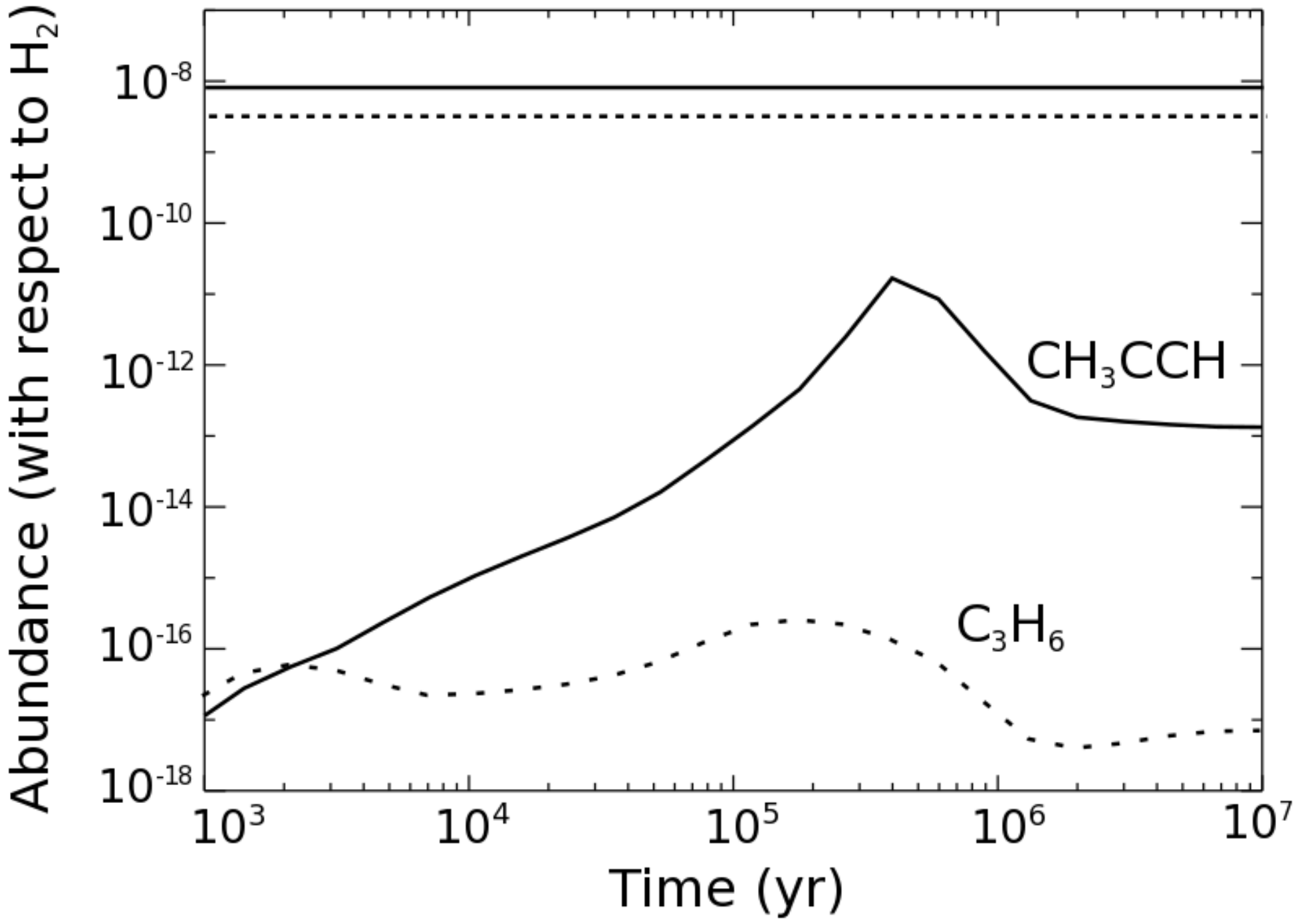

Figure 1

Abundances of methylacetylene (continuous line) and propene (dashed line) as a function of time predicted by our models with $n(H_2) = 2 \cdot 10^4$ $cm^{-3}$ using a model with gas phase reactions only. The horizontal lines represent the abundances observed by Markwick et al. (2002) and Marcelino et al (2007) in TMC-1 (CP): $9.10^{-9}$ for $CH_3CCH$ and $4.10^{-9}$ for $C_3H_6$.

It should be noted that Occhiogrosso et al. recently published an astrochemical model for hydrocarbons, including $CH_3CCH$, in star-forming regions (Occhiogrosso *et al.* 2013). In their study, they used the UMIST 2006 network with some reactions from the online KIDA database and provide in the paper a list of reactions and rate coefficients to be used (in their Table 4). We want to emphasize that many of these values should not be used under astrochemical conditions either because the data are wrong or they cannot be extrapolated to low temperatures and/or densities. We give more explanations below for some systems.

- $C_2H_5 + CH_3CHCH_2 \rightarrow C_3H_5 + C_2H_6$ : in their work Occhiogrosso et al. do not consider any barrier but this reaction has in fact a (high) barrier (Tsang 1991),
- $C_2H + CH_3CHCH_2 \rightarrow C_3H_5 + C_2H_2$ : the rate constant used in Occhiogrosso et al. is coherent with Chataing et al (1998) and Vaktin et al (2001) but the main products are in fact $C_4H_4 + CH_3$ (Woon & Park 2009) and not $C_3H_5 + C_2H_2$ which have been estimated to be equal to 1% (Tsang 1991),
- $C_2H_6 + CH_2CCH \rightarrow C_3H_5 + C_2H_6$ : this reaction appears four times in Table 4 of Occhiogrosso et al., twice with a low "reasonable" estimate of the rate constant value although no studies exist, and twice with a negative barrier leading to k(10K) = 1.2e-9 $cm^3.molecule^{-1}.s^{-1}$ which is clearly incorrect considering the low reactivity of propargyl with closed shell molecules in general and alkanes specifically,
- $C_2H_3 + C_3H_5 \rightarrow CH_3CCH + C_2H_4$ : Occhiogrosso et al. use only one very minor channel of the Tsang (1991) estimation neglecting the main channel,
- $CH_3 + C_3H_5 \rightarrow CH_3CCH + CH_4$ : in this case Occhiogrosso et al. again used only one very minor channel of the Tsang (1991) estimation neglecting the main channel,
- $C_2H + CH_3CHCH_2 \rightarrow CH_3CCH + C_2H_3$ : as above, this is a very minor channel originating from the estimation of Tsang (1991) neglecting the much more recent results of Bouwman et al (2012) leading mainly to the formation of vinylacetylene and methyl and not to methylacetylene,
- $OH + C_3H_5 \rightarrow CH_3CCH + H_2O$ : this is also a minor reaction channel (Tsang 1991) while neglecting the main channel,
- $H + C_3H_5 \rightarrow CH_3CCH + H_2$ and $CH_2CCH_2 + H_2$ : Occhiogrosso et al. used in both cases the high pressure limiting rate constant leading to a large overestimation (Hanning-Lee & Pilling 1992). Moreover, this reaction is supposed to lead only to minor $CH_3CCH + H_2$ production while $CH_2CCH_2 + H_2$ are not produced (Harding et al. 2007)).

The gas phase synthesis of propene is much less efficient than that of methylacetylene. Ionic reactions producing $C_3H_7^+$ involve small fluxes so that the amount of $C_3H_7^+$ generated is too low to efficiently produce $C_3H_6$ through the DR reaction of $C_3H_7^+$. Herbst et al (2010) proposed that $C_3H_7^+$ could be efficiently synthesized through radiative associations: $C_3H_5^+ + H_2 \rightarrow C_3H_7^+ + h\nu$ (2). They performed theoretical calculations leading to a high rate constant for this reaction, in disagreement with (McEwan et al. 1999). Lin et al (2013) also showed that the radiative association leading to protonated propene was slow, whichever $C_3H_5^+$ isomer was considered, so that this process in unimportant in the cold interstellar medium. In a similar manner to methylacetylene, the protonation acts as an overall loss of $C_3H_6$ as the DR of $C_3H_7^+$ does not lead back to $C_3H_6$ with 100% efficiency. Alternative pathways could have been the $CH_5^+ + C_2H_4$ or the $C_2H_5^+ + CH_4$ reactions. However the $CH_5^+ + C_2H_4$ reaction leads to $C_2H_5^+$ + $CH_4$, and not to $C_3H_7^+ + H_2$ (Anicich 2003), and the $C_2H_5^+ + CH_4 \rightarrow C_3H_7^+ + H_2$ reaction shows a barrier (Hiraoka & Kebarle 1975, Hiraoka & Kebarle 1976, Hiraoka et al. 1993, Collins & O'Malley 1994). As a result, these two reactions are not efficient at the low temperature of dense interstellar clouds. The only known reactions producing $C_3H_7^+$ are $C^+ + C_3H_8$, $CH_3^+ + C_2H_6$, $CH_3^+ + C_3H_8$, $C_2H_4^+ + C_2H_6$, $C_2H_5^+ + C_2H_6$ (Anicich 2003) which all involve small fluxes under dense cloud conditions by pure gas phase models, as $C_2H_6$ and $C_3H_8$ are not produced efficiently by gas phase reactions. With regard to neutral reactions, the CH + $C_2H_6$ reaction produces some propene but it involves small fluxes and the radiative association $H + C_3H_5 \rightarrow C_3H_6 + h\nu$ has a low flux as $C_3H_5$ is essentially formed on interstellar grains. The gas phase formation of propene is so inefficient in our network that the calculated abundance remains below $10^{-15}$ as shown in Fig. 1 (dashed line).

It should be noted that the udfa network (http://udfa.ajmarkwick.net/) (McElroy et al. 2012) considers the older (and probably incorrect, as explained above) rate constant values for reactions (1) and (2) taken from (Herbst et al. 2010) which are expected to overestimate the pure gas phase production of both methylacetylene and propene.

## 4. Surface reactions

As gas phase reactions cannot reproduce the observed abundances for methylacetylene and propene, we consider the possibility of surface reactions. Following the formalism of Hasegawa *et al.* (1992), we consider that species formed in the gas phase can stick on grains. At 10 K, only light species (H and $H_2$) can move on the surface. As only H atoms are considered to be reactive, the main reactions are hydrogenation, apart from the specific Eley-Rideal processes introduced by (Ruaud et al. 2015). As a result, $C_3$ (and also l,c-$C_3H$ and l,c-$C_3H_2$ but at lower levels) is efficiently formed in the gas phase in our model, and can stick on the surface, undergo successive hydrogenations to form closed shell s-$C_3H_4$ molecules (s-$C_3H_4$ represents $C_3H_4$ residing on the grain surface). As, by comparison with the gas phase, these hydrogenation reactions are considered to be barrierless (or with very low barriers) up to s-$C_3H_4$ (Mebel et al.

1998, Nguyen et al. 2001, Mebel & Kaiser 2002, Miller & Klippenstein 2003, Hébrard et al. 2013), this is an efficient route for s-$C_3H_4$ synthesis (both methylacetylene and allene, but only the methylacetylene isomer is considered in this work). Also by comparison with the gas phase, the subsequent hydrogenation step to s-$C_3H_5$ shows a small barrier (Whytock *et al.* 1976, Faravelli et al. 2000, Miller et al. 2008, Vuitton *et al.* 2012) which can be overcome through tunneling. The hydrogenation of s-$C_3H_6$ to s-$C_3H_7$ is similarly affected by a small barrier in the gas phase (Tsang 1991, Seakins et al. 1993). In contrast, the radical hydrogenation steps s-$C_3H_5$ + H and s-$C_3H_7$ + H are barrierless in the gas phase (Hanning-Lee & Pilling 1992, Harding et al. 2007, Hébrard et al. 2013). In this study we neglect most of the H atom abstraction reactions. Indeed, for gas phase H + radical (or carbynes such as c-$C_3H_2$) reactions (H + c,l-$C_3H$, c,l,t-$C_3H_2$, $C_3H_3$, $C_3H_5$, $C_3H_7$), the strength of all C-H bonds prevent H atom abstraction to be competitive with H atom addition as there is without doubt a notable barrier for H atom abstraction whereas H atom addition is barrierless. For the H + $C_3H_4$ reactions there are various theoretical and experimental studies (Wang *et al.* 2000, Miller et al. 2008) showing that indeed tunneling is efficient but H atom addition is strongly favoured at low temperature and high pressure (the high pressure limiting rate is expected to be close to the surface reaction one as the energized intermediate can be quickly relaxed through surface interaction). This is also the case for the H + propene reaction (Miller & Klippenstein 2013). Moreover, the H atom abstraction leading to $H_2$ results in radical formation ($C_3H_3$ and $C_3H_5$) which will lead back to $C_3H_4$ and $C_3H_6$ through subsequent H atom addition (H is the only species moving quickly on the surface at 10 K) which is unlikely to change the result of the model.

The binding energies of methylacetylene on icy surfaces have been measured equal to 2500 K (Kimber *et al.* 2014) and the one for propene should be relatively close.

## 5. Modeling results

The relative abundances with respect to $H_2$ of the various s-$C_3H_x$ compounds (with x = 0 to 8) computed by our chemical model (with the reactions listed in Table I) are shown Fig. 2. The cyclic and linear $C_3H$ and $C_3H_2$ species have the same predicted abundances on the surfaces.

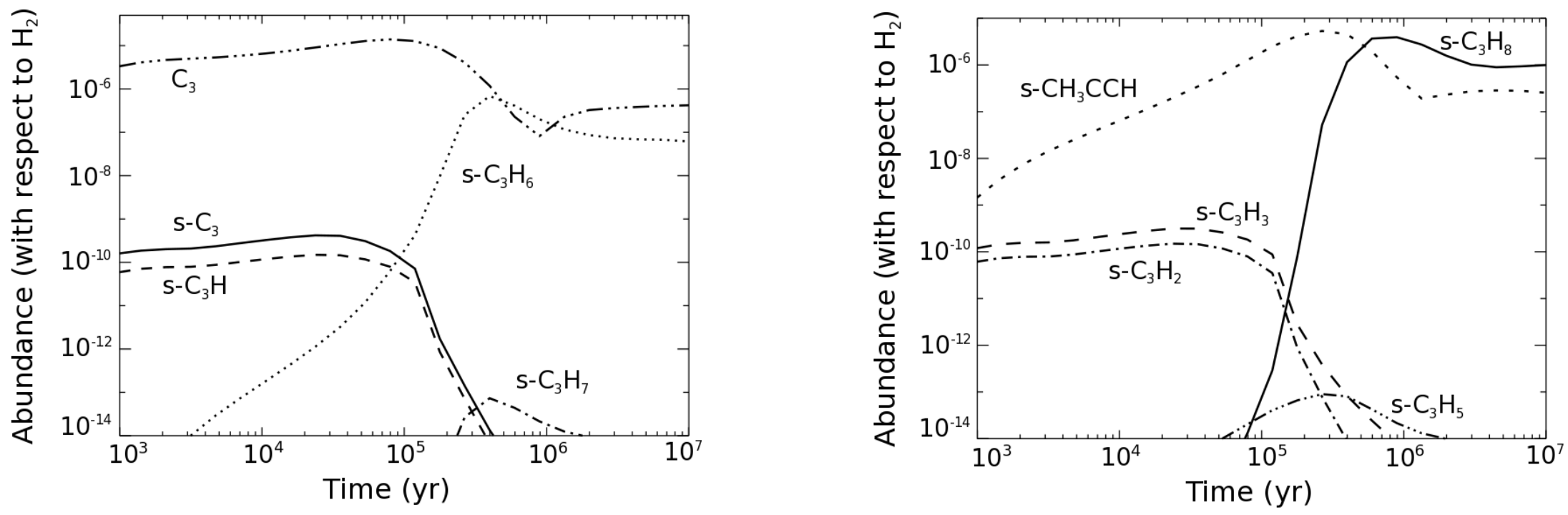

Figure 2

Abundances of $C_3$ in the gas phase and s-$C_3H_{x=0\text{-}8}$ surface species as a function of time, predicted by our model.

The formation of s-$C_3H_4$ in a few $10^4$ yr is explained by the depletion onto ices of $C_3$ produced in the gas phase followed by the quick transformation of s-$C_3$ into s-$C_3H_4$ through barrierless H atom addition. s-$C_3H_4$ is then more slowly converted into s-$C_3H_6$ and s-$C_3H_6$ into s-$C_3H_8$ in a few $10^5$ yr (time scale comparable to dense cloud ages). The smaller peak abundance of s-$C_3H_6$ relative to s-$C_3H_4$ is due to the fact that s-$C_3H_6$ is more easily hydrogenated than s-$C_3H_4$. At $10^6$ yr a large part (20%) of the $C_3$ produced in the gas phase is transformed into s-$C_3H_8$ which represents 1 % of the initial carbon, the main reservoir of carbon atoms being CO and s-CO. As $C_3H_8$ is relatively non reactive on ices at 10 K, further evolution is mainly driven by photodissociationprocesses which are almost unknown for species in ices. Then the composition of the ice given by the model is highly hypothetical for times greater than $10^6$ yr although the gas phase $CH_3CCH$ and $C_3H_6$ abundances at times relevant to dense molecular clouds (around few $10^5$ years) are unaffected.

The gas phase $C_3H_4$ and $C_3H_6$ abundances (relative to $H_2$) are shown in Fig. 3 together with the observed abundances in cold cores.

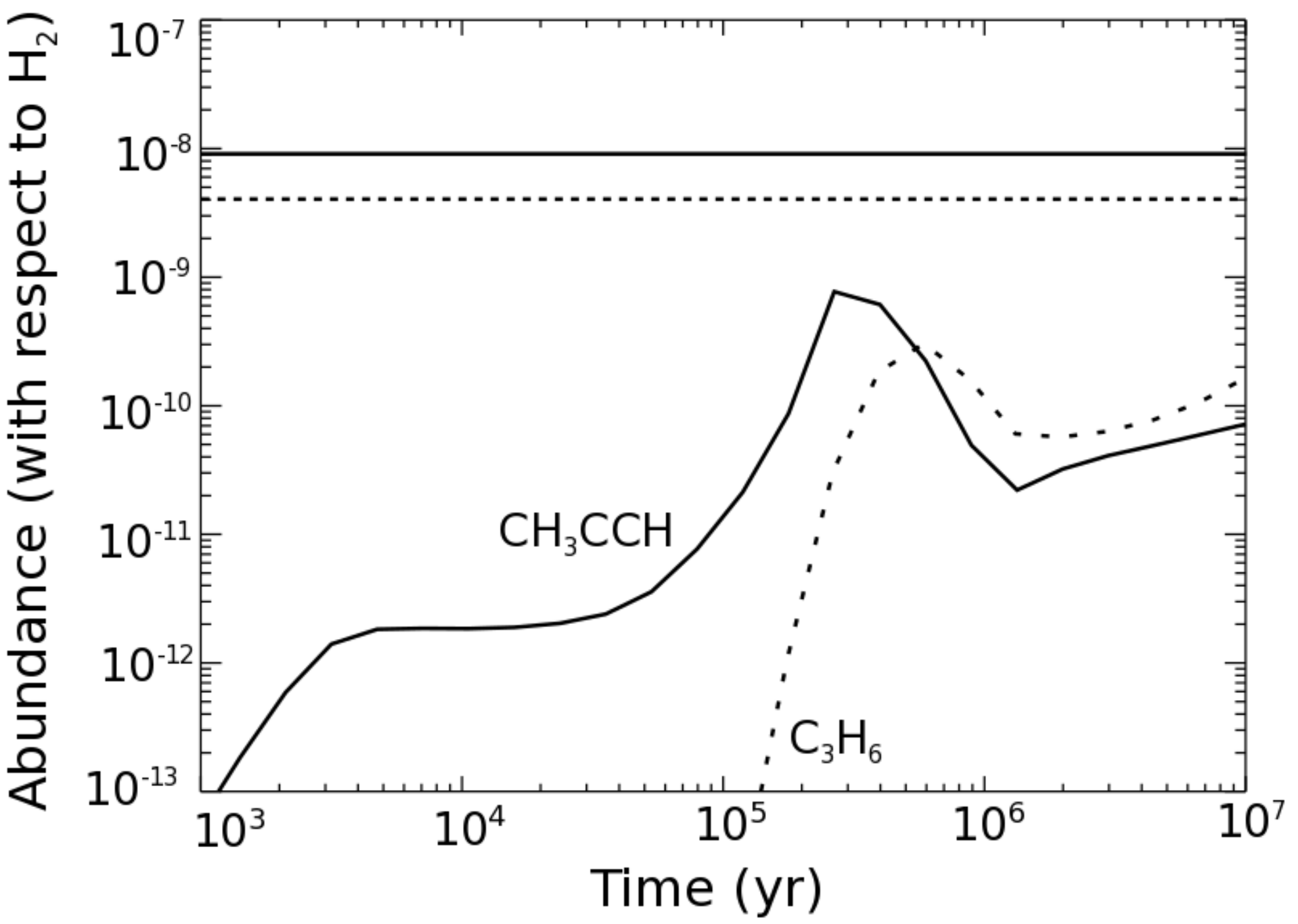

Figure 3

Abundances of methylacetylene (continuous line) and propene (dashed line) species as a function of time predicted by our model. The horizontal lines represent the abundances observed by Markwick et al. (2002) and Marcelino et al (2007) in TMC-1 (CP): $9.10^{-9}$ for $CH_3CCH$ and $4.10^{-9}$ for $C_3H_6$.

The agreement between observations and calculations is much better than with the model considering gas phase reactions only (see Fig. 3). The predicted abundances in the gas-phase are however still an order of magnitude smaller than the observational values in TMC-1 (CP) at typical cloud ages (around few $10^5$ years). Considering the large amount of s-$C_3H_4$ and s-$C_3H_6$ on the grains, more efficient non-thermal desorption, such as chemical desorption for instance, improves the agreement with the observations. With a higher chemical desorption efficiency, from 1% up to 10 %, we obtain a much better agreement between calculations and observations for $CH_3CCH$ and $C_3H_6$, but also for $CH_3OH$, $H_2CO$ and NH species which are partly or mostly formed on grains (Ruaud et al. 2015). However, this results in less good agreement for $NH_3$ which becomes over-abundant. Chemical desorption is a complex mechanism that very likely depends on the species involved. Very recent work by Minissale et al. (Minissale *et al.* 2016) proposes a semi empirical model to estimate the efficiency of chemical desorption for particular species. From this model, using the binding energy of methylacetylene on amorphous water? ice measured by (Kimber et al. 2014) (equal to 2500 K) for methylacetylene and propene, leads to about 1 % of chemical desorption. This efficiency is highly dependent on the binding energy as well as the semi-empirical parameter ε of (Minissale et al. 2016) which corresponds to the fraction of kinetic energy retained by the product colliding with the surface. It should be noted that other non-thermal desorption mechanisms, such as cosmic-ray explosive desorption (Ivlev *et al.* 2015) may play a role. More accurate predictions of species formed on interstellar grains can only be provided by more detailed models of these surfaces, as well as through precise measurements (or calculations) of the efficiency of the chemical desorption mechanism in interstellar ice analogs. In our nominal model we keep the fraction of desorption for the newly formed species on grains equal to 1% which seems a reasonable value considering the latest work from Minissale et al (2016).

## 5. The $C_3$ + O reaction

The chemistry of $C_3H_4$ and $C_3H_6$ on grains is triggered by the depletion of gas phase $C_3H_x$ species (where x can be zero). Then the amount of $C_3H_x$ in the gas phase controls the grain production of $C_3H_4$ and $C_3H_6$ (and therefore the amount of $C_3H_4$ and $C_3H_6$ released back into the gas phase through reactive desorption). In our network, the most abundant $C_3H_x$ compound produced in the gas phase is $C_3$ (by far), reaching a maximum abundance of $10^{-5}$·[$H_2$] at $10^5$ years so that efficient $C_3H_4$ and $C_3H_6$ production on grains is directly linked to the large gas phase $C_3$ abundance. $C_3$ is abundant in the gas phase as a result of various efficient neutral pathways producing $C_3$ (among them the C + $C_2H$ and C + $C_2H_2$ reactions) with very few destruction mechanisms. Indeed, $C_3$ has a low reactivity with abundant species in molecular clouds such as O, H, N, CO, $CH_4$, $C_2H_2$ (Woon & Herbst 1996, Nelson *et al.* 1982, Nelson *et al.* 1981, Mebel *et al.* 2007, Guo *et al.* 2007, Li *et al.* 2005, Szczepanski *et al.* 2004). $C_3$ reacts without a barrier with C atoms leading to $C_4$ through radiative association (Wakelam *et al.* 2009), with highly reactive radicals such as CH and $C_2H$ species (Hébrard et al. 2013) these reactions involving small fluxes, and with $H_3^+$, $HCO^+$ and $HCNH^+$ through proton transfer. As $C_4$ leads mainly back to $C_3$ through reactions with C, O and N atoms (Wakelam et al. 2009), the only efficient reactions for $C_3$ loss are with $H_3^+$, $HCO^+$ and $HCNH^+$ and through depletion onto grains. The low reactivity of $C_3$ allows it to reach high abundance levels, which is at the origin of the rich $C_3H_x$ chemistry on grains. To estimate the importance of the O + $C_3$ reaction, which has been studied only theoretically by (Woon & Herbst 1996) who found a small barrier in the entrance valley, we performed a run considering no barrier for this reaction and a rate constant equal to $2.0 \cdot 10^{-10}$ $cm^3$ $s^{-1}$ $molecule^{-1}$ (close to the capture rate constant). The results for the abundances of $C_3H_4$ and $C_3H_6$ in the gas phase are shown in Fig. 4.

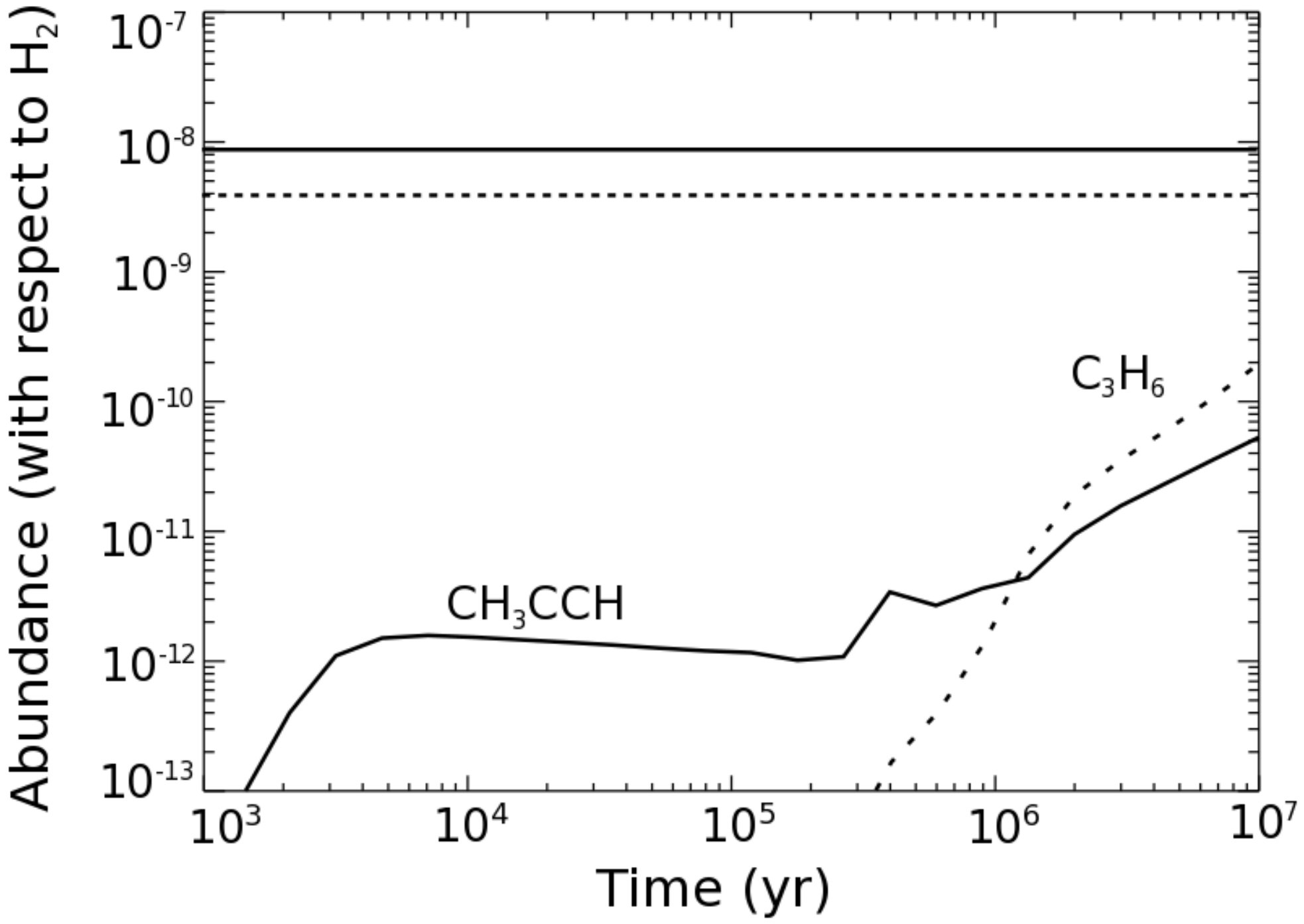

Figure 4

Abundances of gas phase methylacetylene (continuous line) and propene (dashed line) as a function of time predicted by our model considering no barrier for the O + $C_3$ reaction. The horizontal lines represent the abundances observed by Markwick et al. (2002) and Marcelino et al (2007) in TMC-1 (CP): $9.10^{-9}$ for $CH_3CCH$ and $4.10^{-9}$ for $C_3H_6$.

It is clear that the simulated abundances cannot reproduce the observed ones for typical cloud ages of a few $10^5$ years. The abundances of these species on the surfaces are also strongly decreased, yielding values below the gas-phase ones. It is worth noting that the absence of barrier for the O + $C_3$ reaction affects various large species in the gas phase, namely $C_3H_{x=0-6}$, $C_3O$, $C_3S$, $C_3N$, $HC_3N$ and $C_4H$. For almost all these species, the absence of barrier for the O + $C_3$ reaction leads to large disagreements between calculated abundances and observations in the current state of the model. It appears likely that the calculations of Woon & Herbst (1996) for the O + $C_3$ reaction are reliable and should be used in chemical networks.

We have also looked to the $C_3$ + H → $C_3H$ reaction which is calculated to be barrierless (Mebel & Kaiser 2002, Hébrard et al. 2013) but may show a small barrier considering the theoretical uncertainties. In the gas phase, even without a barrier, the radiative association rate is estimated to be below $10^{-14}$ $cm^3$ $s^{-1}$ at 10 K (Hébrard et al. 2013) so this process has a negligible flux. For the surface reaction, s-$C_3$ + s-H → s-$C_3H$, the presence of a small barrier has very little effect on s-$CH_3CCH$ and s-$C_3H_6$ abundances. So, in contrast to the O + $C_3$ reaction, the H + $C_3$ reaction is not a critical one in dense molecular clouds.

## 6. Conclusion

We have shown, using our gas-grain chemical model NAUTILUS, that surface reactions may explain the observed abundances of methylacetylene and propene in cold dense molecular clouds. Based on the recent calculations by Lin et al (2013) and the current review of gas-phase reactions, gas-phase reactions are unlikely to produce enough of these species. These results are in good agreement with the suggested scenario by Miettinen et al (2006) who found a gas-phase methylacetylene abundance which increases with the temperature in star forming regions, suggesting its formation on ice surfaces followed by thermal desorption. We also highlight the critical role of the O + $C_3$ reaction which is likely to possess a substantial barrier, in good agreement with the calculations of Woon & Herbst (1996) leading to a high $C_3$ abundance in dense molecular clouds.

JCL and KMH thank the French CNRS/INSU program PCMI for their partial support of this work. VW researches are funded by the ERC Starting Grant (3DICE, grant agreement 336474).